\documentclass[12pt]{iopart}
\usepackage{graphicx}
\usepackage{braket}
\usepackage{amssymb}
\newcommand{\av}[1]{\langle\, #1\,\rangle}
\begin{document}

\title{From ergodic to non-ergodic chaos in Rosenzweig-Porter model}

\author{M. Pino$^\dagger$, J. Tabanera$^\ddag$, P. Serna$^*$}

\address{$\dagger$ Institute of Fundamental Physics IFF-CSIC, Calle Serrano 113b, Madrid 28006, Spain\\
$\ddag$ Department of Structure of Matter, Thermal Physics and Electrodynamics, Universidad Complutense de Madrid, 
Pl. de las Ciencias 1. 28040 Madrid, Spain\\
$^*$ Laboratoire de Physique de l'\'Ecole Normale Sup\'erieure, ENS, Universit\'e PSL, CNRS, Sorbonne Universit\'e,
Universit\'e Paris-Diderot, Sorbonne Paris Cit\'e, Paris, France.}
\ead{mpg@iff.csic.es}

\begin{abstract}

The Rosenzweig-Porter model is a one-parameter family of random matrices with three different phases: ergodic, extended non-ergodic and localized.  We characterize numerically each of these phases and the transitions between them. We focus on several quantities that exhibit non-analytical behaviour and show  that they obey the scaling hypothesis. Based on this, we argue that non-ergodic chaotic  and ergodic regimes are separated by a continuous phase transition, similarly to the transition between non-ergodic chaotic and localized phases.

\end{abstract}

\section{Introduction}

The conditions under which quantum systems exhibit chaos or thermalization are not yet fully understood. The Bohigas–-Giannoni–-Schmit conjecture and the thermalization eigenstates hypothesis relate the Hamiltonian of complex quantum systems, complex enough to display a strong form of chaos, with Gaussian ensembles of random matrices\ \cite{Bohigas1984,Sr1999,Dalessio16}.  However, these ensembles fail to describe non-thermal quantum dynamics, as the one that occurs in  many-body localized Hamiltonians\ \cite{Basko2006}.

We study a one-parameter family of random matrices known as Rosenzweig-Porter (RP) model\ \cite{rosenzweig1960,altland1997perturbation}. The behaviour of this model agrees well with the predictions of Gaussian ensembles when the parameter is small, we call this regime ergodic, and it shows properties of integrable Hamiltonians when it is large. We refer to the latter regime as localized since each of its wavefunctions is confined in a small region of the Hilbert space. For intermediate values of the parameter the wavefunctions are extended but non-ergodic\ \cite{Kravtsov2015}, a behaviour which was first discussed for interacting electrons in quantum dots\ \cite{altshuler1997quasiparticle}. The system displays chaos in this regime, as we will see, but this chaos is not strong enough to reproduce the full behaviour of Gaussian ensembles. 

Several authors have investigated the non-ergodic extended wavefunctions  with different techniques\ \cite{monthus2017multifractality,truong2016eigenvectors,von2017non,amini2017spread,de2018survival, Bera2018,de2019survival}. In  particular, for the RP model an analytical method based on Green's function \cite{Ab1979} has been shown to be able to distinguish the three different regimes\cite{facoetti2016non}. Building upon similar ideas, a general method to obtain the properties of the non-ergodic metallic phase for any model was also published \cite{altshuler2016multifractal}.

There are many-body systems with some similarities to the problem of a single particle in a space with large dimension\ \cite{shukla2016localization}. This is the case for many-body localization\ \cite{Basko2006} and for some relevant models used in quantum annealing\ \cite{altshuler2010anderson,knysh2010relevance,pino2018quantum}. Those systems are quite difficult to analyze and one can try, as a first approximation, to model them with a random Hamiltonian with large connectivity, as the RP model\ \cite{shukla2016localization}. The results and techniques presented here may help in the study of those  many-body quantum systems.

The most interesting property of the RP model is the existence of a finite region of non-ergodic extended states. Such a region appears at the metallic side of the many-body localization transition in  an array of Josephson junctions\ \cite{Pino15, pino2017, thudiyangal2018dynamical, Mithun2018, Mithun2019}. This non-ergodic behaviour has also been analyzed in other quantum systems\ \cite{torres2017extended, berkovits2017signature, lindinger2018many, roy2018multifractality, faoro2018non} and it has been suggested that it can play an important role for quantum information\  \cite{kechedzhi2018efficient}. On top of that, there are several teams that have obtained a sub-diffusive but ergodic behaviour in one-dimensional spin chains at the metallic side of the many-body localization transition\ \cite{luitz2017ergodic, Argarwal2015, Lev2015, Luitz2016, Argarwal2015, vznidarivc2016diffusive, Biroli2017}. 

Non-ergodic extended states also occur for models of a single-particle in a lattice with disorder. Its existence is well established for the Anderson model in three dimensions, but the  region where they appear is a single point in the parameter space\ \cite{Evers2008}. The situation seems to be different for the Anderson model in a random regular graph, where a non-ergodic  extended phase has been found\ \cite{Deluca2014, Al2016, kravtsov2018non}. In this case, the transition between non-ergodic and ergodic phases, different than the Anderson localization transition, has been shown to be of first order\ \cite{kravtsov2018non,altshuler2016multifractal}. 

The whole picture of non-ergodic wavefunctions in the metallic side of the Anderson transition for random regular graph has been challenged in references\ \cite{garcia2017scaling, Tikhonov2016, garcia2019two}. There, it was shown both that the metallic wavefunctions are ergodic in the thermodynamic limit and that non-ergodicity is only restricted to small length scales. Specifically, a finite size scaling approach \cite{garcia2017scaling} for this transition has been developed which is consistent with a single localization transition and a crossover from non-ergodicity to ergodicity in the metallic phase. The method used there utilizes different scaling laws  at each side of the localization transition.

Here, we will focus on the numerical characterization of the non-ergodic extended regime and in its phase transitions. In our approach, we focus on quantities that harbour non-analytical behaviour at those transitions in the limit of large matrices size. These discontinuities are smoothed in finite systems, but via finite size scaling we show that they can be explained with the existence of a single critical length that diverges at the transition. We will obtain scaling collapse with a single scaling function, in contrast to reference \cite{garcia2017scaling}, at both sides of the transitions. The structure of the paper is as follows. In Sec.\ \ref{sec:intro} we explain the model and give a qualitative picture of the different phases. Next, in Sec.\ \ref{sec:ncha}, we present several quantities based on the probability distribution of eigenstates and eigenfunctions that will be used to  characterize each phase. In Sec.\ \ref{sec:cha_fss}, we will show the scaling collapse of different quantities for each one of the phase transitions. The last section contains a summary of the results.

\section{Rosenzweig-Porter model}\label{sec:intro}

The RP model is a one-parameter family of random matrices.  Their matrix elements are all distributed following a Gaussian distribution with zero mean, but  the variances are different for diagonal and non-diagonal elements.  We will work with the orthogonal version of the RP model, {\it i.e.}, real symmetric matrices. The entries of this model are distributed following: 
\begin{equation}
 \av{H_{ii}^2} = 1  \hspace{4cm}  \av{H_{ij}^2} = \frac{ N^{-\gamma}}{2},\label{eq:rp}
\end{equation}
where $N$ is the size of the Hilbert space and $\gamma>0$. The properties of the eigenstates at the middle of the spectrum will be analyzed as a function of the parameter $\gamma$. 

The RP model has large connectivity, all the matrix elements are different from zero, although the off-diagonal elements are pretty small for $\gamma\gg 1$. It has been argued that the perturbation series, taking the non-diagonal elements as the perturbation, absolutely converges for $\gamma>2$\ \cite{Kravtsov2015}. Consequently, eigenstates are localized around a few state of the computational bases similarly to an Anderson insulator\ \cite{Kravtsov2015,An1958}. This phase presents typical properties of  integrable systems like a Poisson level distribution and will be called localized.

The perturbation theory diverges for $\gamma<1$ and the eigenstates display metallic behaviour. This regime will be called ergodic as it is well described by the Gaussian Orthogonal Ensemble (GOE) of random matrices.

The most interesting region of parameters is $1<\gamma<2$, where the wavefunctions change with the size of the matrix following a multifractal scaling\ \cite{Kravtsov2015}. This implies that the wavefunctions has a support set, $C$, in the computational basis which is similar to a fractal $C\sim N^{D}$ with $D<1$. In consequence, the number of vectors in the computational basis that overlap with the wavefunction is large but it remains a small fraction of the total. We call this regime extended non-ergodic following  reference\ \cite{Kravtsov2015}. See references \cite{Kravtsov2015,facoetti2016non, bogomolny2018eigenfunction} for an exhaustive analytical treatment of this regime.

\section{Numerical characterization of the RP phases}\label{sec:ncha}

This section is devoted to the numerical characterization of the different phases of the RP model. We have  generated and diagonalized around $10^3$ random matrices following the RP distribution in equation\ \ref{eq:rp} for each value of $\gamma$ and matrix dimension $N$.  Then, we have computed quantities related to the level-spacing (explained in subsection\ \ref{sec:cha_energies}) and to the eigenstates  (defined in subsection\ \ref{sec:cha_eigenstates}).

\subsection{Characterization via eigenenergies}\label{sec:cha_energies}
We first look at several quantities that provide insight into the distribution of the  spectrum of energies in RP model. We are interested on the distribution of adjacent level-spacing $r=\av{\min(\delta^n-\delta^{n+1})/\max(\delta^n-\delta^{n+1})}$, where $\delta^n=E_{n+1}-E_{n}$ is the $n$-th level spacing and the brackets $\av{\dots}$ means average over the matrix probability distribution. Its value is $r\approx0.5307$ for level-spacing distributed as GOE and $r\approx 0.3863$  for Poisson\ \cite{Atas2013}. 

\begin{figure}[t!]
\begin{centering}
\includegraphics[width=1.\columnwidth]{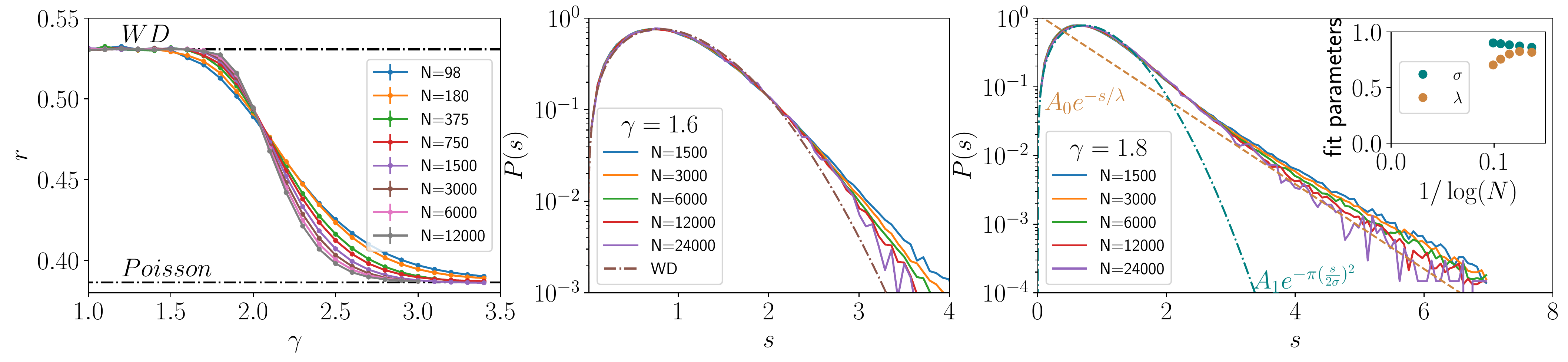}
\par\end{centering}
\vspace{0.1cm}
\caption{Left panel contains the average of ratio between adjacent levels $r=\min(\delta^n-\delta^{n+1})/\max(\delta^n-\delta^{n+1})$, where $\delta_n=E_n-E_{n-1}$ is the $n$-th level spacing, as a function of $\gamma$. The normalized distribution of level-spacing is plotted as a function of  $s=\delta/\av{\delta}$ for $\gamma=1.6$ (center) and $\gamma=1.8$ (right). The semi-dashed curve in the center plot corresponds to the Wigner-Dyson distribution $P_{WD}(s)=\pi s/2\exp{(-\pi s^2/4)}$. On the right panel, the dashed and semi-dashed curves represent a fitting curve of the numerical data for the largest size $L=24000$ to $P(s)=A_0 \exp{\left(-s/\lambda\right)}$ (in $2<s<7$) and  $P(s)=A_1\ s\ \exp{\left[-\pi (s/2\sigma)^2\right]}$ (in $0<s<2$), respectively. The free parameters of these fittings are $\lambda$, $\sigma$ $A_0$ and $A_1$. Repeating the same procedure for all the sizes produces a set of parameters $\sigma$ and $\lambda$ for different sizes, which are plotted as a function of $1/\log(N)$ in the inset.
}\label{Fig_ls}
\end{figure}

The value of $r$ as a function of $\gamma$ appears in the left panel of Fig.\ \ref{Fig_ls}.  There is a point at which the curves for different sizes cross. This crossing marks the presence of a non-analytic point when the system size goes to the thermodynamic limit. Thermodynamic limit means here that the matrix dimension goes to infinity. In the following section, we will see that this kind of non-analytical behaviour appears in several other quantities. The results for $r$ are compatible with a phase transitions at $\gamma=2$, however there is no sign of a second transition at $\gamma=1$. Indeed, we cannot distinguish the value of $r$ in the ergodic and non-ergodic extended phases once finite size effects are taking into account.

We have also analyzed the full distribution of level-spacing. In Fig. \ref{Fig_ls}, this distribution is computed with different system sizes for $\gamma =1.6$ (center) and $\gamma=1.8$ (right). It is clear that both cases show level repulsion, which means $P(s)\sim s^p$ for small $s$ and $p>0$. The presence of level repulsion evidences that chaos is already present in the non-ergodic extended regime. 

The level spacing distribution for $\gamma=1.8$ exhibits significant differences with Wigner-Dyson,  $P_{WD}(s)=\pi s/2\exp{(-\pi s^2/4)}$. To quantify their dependence with system size, we fit the distributions to the theoretical limiting distributions, Wigner-Dyson and Poisson (with arbitrary parameters). For Wigner-Dyson, we fit each curve in the region $0<s<2$ to the function $P(s)=A_1\ s\ \exp{\left[-\pi (s/2\sigma)^2\right]}$, where $A_1$ and $\sigma$ are free parameters. For Poisson, the region $2<s<7$ is used to fit each curve to the function $P(s)=A_0 \exp{\left(-s/\lambda\right)}$, where $\lambda$ and $A_1$ are free parameters. The fitted parameters $\sigma$ and $\lambda$ are represented as a function of $\log(N)$ in the inset of right panel of  Fig.\ \ref{Fig_ls}. Although far from the limits, there is a clear trend that takes $\sigma\rightarrow1 $ and $\lambda\rightarrow0$. Thus, finite size effects are large in the level spacing for $\gamma=1.8$ and we cannot exclude that the Wigner-Dyson law appears in the thermodynamic limit. 

This is also supported by the fact that our numerical data are fully consistent with r-statistic having the same value in the non-ergodic extended regime and in GOE. We have found similar results for other values of $\gamma$ inside the non-ergodic extended regime.

In summary, the numerical results we present here points to the statistic of energy levels changing from Poisson, $\gamma>2$, to Wigner-Dyson at $\gamma<2$, as discussed in\ \cite{facoetti2016non} and also in \cite{kunz1998, brezin1996, altland1997perturbation, shukla2000} for the Gaussian unitary RP model. At the same time, a non-analytical behaviour is clearly seen in the r-statistic in the thermodynamic limit. Specifically, the value of $r=0.5307$ does not necessarily imply GOE behaviour as it seems to happen for non-ergodic eigenstates in the RP model.  This should be taken into account in the field of many-body localization, where this quantity has been extensively used to characterize the metal-insulator transition\ \cite{Oganesyan2007}. In the following, we will analyze other quantities based on the  distribution of eigenstates in order to access the second transition and characterize numerically the non-ergodic phase.

\subsection{Characterization via eigenstates}\label{sec:cha_eigenstates}

We focus on the participation entropy and Kullback-Liebler divergences, all of them calculated in the computational basis\ \cite{Luitz15}. We use $\psi_a^n(i) = \braket{i|\psi_a^n}$ to denote the projection of the n-th eigenvector $\ket{\psi_a^n}$ of realization $H_a$ on  a vector $\ket{i}$ of this basis. The participation entropy is:
\begin{equation}\label{eq:S}
S = \braket{\sum_{i=1}^{N}   |\psi_a^n(i)|^2  \log{  \left(|\psi_a^n(i)|^2\right) }  }, 
\end{equation}
where $\braket{\dots}$ is an average over the probability distribution of the matrices and $N$ is the dimension of the Hilbert space.

Two versions of the KL divergence are considered. We use $KL_1$ and $KL_2$ for the Kullback-Liebler divergence of two eigenstates of the same  and different realization, respectively. That is:
\begin{eqnarray*}
 KL_1 = \braket{\sum_{i=1}^{N} |\psi_a^n(i)|^2\log{\left(\frac{ |\psi_a^n(i)|^2}{|\psi_a^m(i)|^2}\right)}}, \\
 KL_2 = \braket{\sum_{i=1}^{N} |\psi_a^n(i)|^2\log{\left(\frac{ |\psi_a^n(i)|^2}{|\psi_b^m(i)|^2}\right)}}. 
\end{eqnarray*}
where $n$ and $m$ are different but close integer numbers so the two eigenvectors are at the same energy density.

Intuitively, the KL divergence estimates  the overlap between two vectors:  it is large if the wave-functions are spread in different regions of the Hilbert space and is of order one for wavefunctions with the same support set. Level repulsion implies that nearby eigenstates hybridizes, so they occupy similar regions in Hilbert space giving a finite $KL_1$. However, $KL_1$ would diverge in the absence of this repulsion. We have seen in figure\ \ref{Fig_ls}  that level repulsion appears for $\gamma\leq 2$, so we expect a divergence of $KL_1$ at $\gamma>2$. One of the features of non-ergodic states is that they do not expand over the whole Hilbert space, and the definition of $KL_2$ provide insight into these states. The reason is that there cannot be any correlation between the support sets of two wavefunctions for different samples at the middle of the spectrum. Due to this independence, the chances are that the support set of two non-ergodic wavefunctions do not fully overlap. From this it follows the divergence of $KL_2$. This happens when $\gamma>1$. We will see that the location of the phase transitions at $\gamma=2$ and $\gamma=1$ can be precisely estimated using the divergences of $KL_1$ and $KL_2$, respectively. In the following, we use the multifractal ansatz to formalize this intuition.

\begin{figure}[t!]
\begin{centering}
\includegraphics[width=0.9\columnwidth]{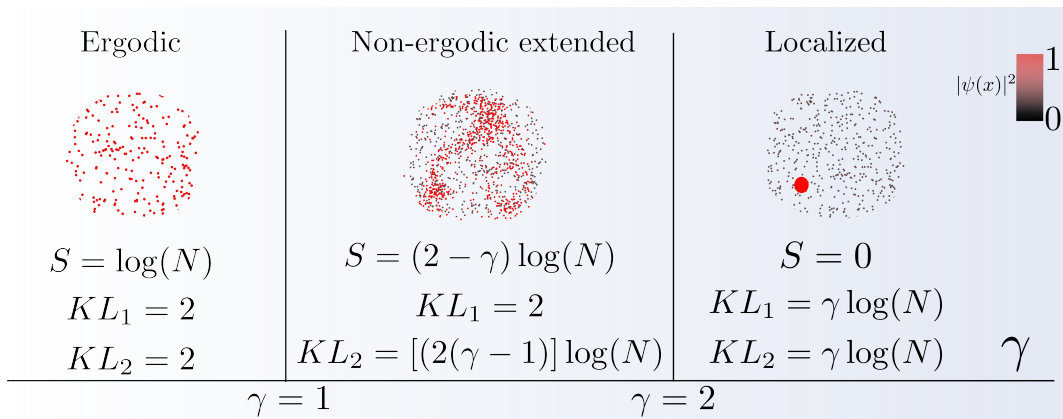}
\par\end{centering}
\vspace{0.2cm}
\caption{Schematic representation of the eigenstates of the RP model at each of its phases and a summary of the values of participation entropy $S$, and KL divergences to leading order. Each of the dots represents a vector in the computational basis. The amplitude of the wavefunction at each of this vectors goes from 0 (black) to 1 (red). It should be emphasized that the RP Hamiltonian contains an all to all connection so there is not any notion of dimensionality. At $\gamma<1$, the eigenstates follow the behaviour of GOE while they obey a multifractal scaling in the non-ergodic extended regime, $1<\gamma<2$. Deep in the localized phase, $\gamma \gg 2$, the eigenstates are Anderson localized. 
}\label{Fig:sc}
\end{figure}

The values of the entropy and KL divergences can be found for $\gamma<1$ by doing Gaussian integrals since this limit is well described by GOE. They are $S =  \log(\frac{2 e^{\gamma-2}}{N})$, where we used momentarily $\gamma$ to denote Euler constant,  and $KL_1 = KL_2 =2$.  The wavefunction is concentrated in a small region of the Hilbert space in the localized phase $\gamma>2$\ \cite{Kravtsov2015}. Second order perturbation theory, taking non-diagonal elements as the perturbation, describes the wavefunctions in the strongly localized limit, $\gamma\gg 2$. In this limit, participation entropy is $S=0$ and both Kullback-Liebler divergences are $KL=\gamma \log(N)$.
 
The wavefunctions are multifractal in the intermediate regime $1<\gamma<2$\ \cite{Kravtsov2015}.  Multifractality can be characterized via the fractal dimensions $D_q$, defined as $I_q = \av{\sum_{i}|\psi(i)|^{2q}} \sim N^{D_q(1-q)}$, or by the spectrum of singularities $f(\alpha)$\ \cite{halsey1986fractal}.  In fact, $f(\alpha)$ is the Legendre transform of $D_q$ so $D_q=[q\alpha_q -f(\alpha_q)]/(q-1)$ and $f^\prime(\alpha_q)=q$. One can show by using this multifractal ansatz and applying the steepest descent approximation that: 
\begin{eqnarray}
S &=& \alpha_1\log{N}+O(1),\label{eq:S}    \\
KL_2 &=& \left(\alpha_0 - \alpha_1\right)\log{N}+O(1).\label{eq:KL2m}
\end{eqnarray}
being $\alpha_1 = D_1$ \cite{monthus2007multifractal}. The inequality $\alpha_0\geq\alpha_1$ holds due to the convexity of the spectrum of singularities $f(\alpha)$.  For the analysis of numerical data it is useful to consider the quantity $s=S/\ln(N)$.

\begin{figure}[t!]
\begin{centering}
\includegraphics[width=1.\columnwidth]{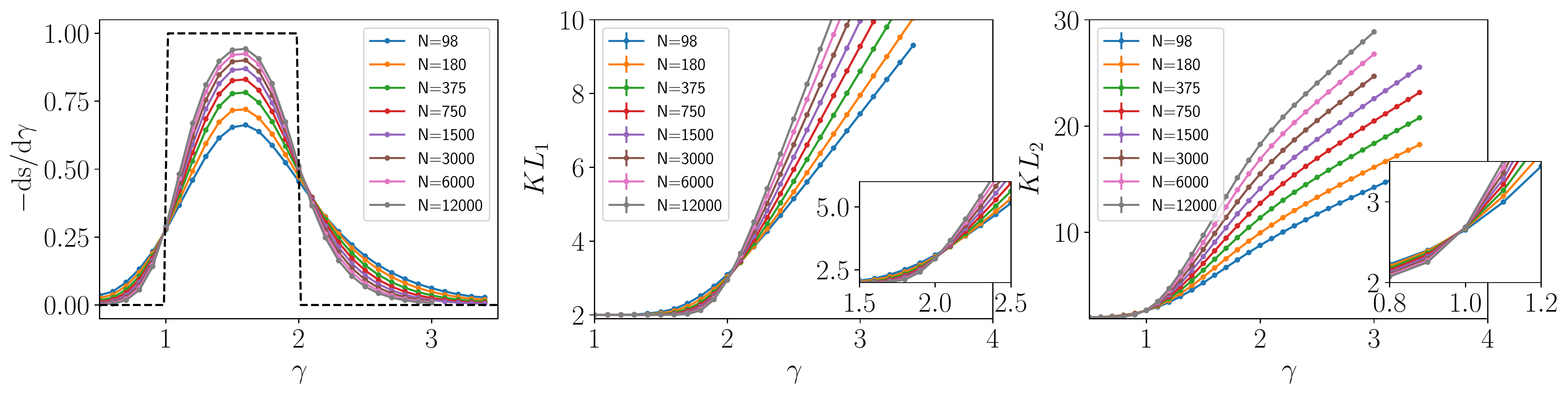}
\par\end{centering}
\vspace{0.2cm}
\caption{ Left panel contains  the derivative $-\frac{ds}{d\gamma}$, where $s=S/\ln(N)$, as a function of $\gamma$. The dashed line represents this quantity in the limit of $N\rightarrow \infty$.  The quantity $KL_1$ is plotted as a function of $\gamma$ in the central panel and its inset zooms in the region where  curves for different sizes cross. In the right panel, $KL_2$ is shown as a function of $\gamma$.  The inset zooms in again into the crossing point. For all the data, each color represent a value of matrix dimension $N$.
}\label{Fig_sev}
\end{figure}

The spectrum of fractal dimensions for the extended non-ergodic regime of RP model can be computed as in reference\ \cite{Kravtsov2015}.  The wavefunctions scale with fractal dimensions $D_q = 2-\gamma$ for $q>1/2$ and $D_q=(q\gamma-1)/(q-1)$ for $q<1/2$.  A fractal is the particular case when $D_q=D$ for all $q$, a different scenario to what happens here. The wavefunctions are thus multifractals although the moments $I_q$  for $q\geq 1/2$ behave like the case of a pure fractal. The value of $\alpha_0=\gamma$ and of $\alpha_1=2-\gamma$ can be used to compute $S$ and $KL_2$ from equations\ \ref{eq:S} and \ref{eq:KL2m}.  We have summarized the different values of entropy and KL divergences in Fig.\ \ref{Fig:sc}. The quantity $KL_1$ in the non-ergodic regime is harder to evaluate analytically than $KL_2$, although it is clear in Fig.\ \ref{Fig_sev} that $KL_1=2$. We have shown that the distribution of level spacing is Wigner-Dyson. This implies that wavefunctions  that are close in energy hybridize as GOE eigenvectors, and we would expect a $KL_1\approx 2$. In the insulator regime, fractal dimensions are zero $D_q=0$ only for $q>1/\gamma$ and non-zero for $0<D_q<1/2$\ \cite{Kravtsov2015}. Anderson localization, $D_q=0$ for all $q>0$, is fully recovered when $\gamma\gg 1$.

In  Fig.\ \ref{Fig_sev}, we can see the results for $S$, $KL_1$ and $KL_2$ computed for $5000$ samples and several sizes.  The derivative $-\frac{ds}{d\gamma}$, where $s=S/\ln(N)$, appears as function of $\gamma$ in the left panel. There are two points at which the curves for different sizes cross. These crossing points evidence that a discontinuity develops as the system size is increased. As usual, the abrupt transition only happens in the thermodynamic limit, meaning infinite matrix size. We can infer from there that the derivative of the entropy is a good quantity to characterize both transitions, at $\gamma=1$ and $\gamma=2$. The dashed line shows the limit of $-\frac{ds}{d\gamma}$ for $N\rightarrow \infty$.

$KL_1$ and $KL_2$ divergences are shown in the center and right panel of Fig.\ \ref{Fig_sev}, respectively. The crossing of curves for different sizes in $KL_1$ are consistent with a transition at $\gamma=2$, while the crossing of curves for $KL_2$ happens at $\gamma=1$ where the non-ergodic to ergodic extended transition occurs. This fits well with the analysis made before.  

The results in Fig.\ \ref{Fig_sev} show that the entropy and the Kullback-Liebler divergences are good quantities to locate both of the critical points in the RP model. In the next section, we characterize their behaviour at the two critical points seeking their scaling collapse, via finite size scaling.

\section{Scaling hypothesis}\label{sec:cha_fss}

Some measurable quantities may display a singularity at a critical point, albeit only in the thermodynamic limit. However, due to the restrictions of numeric techniques we only have access to systems of finite size. A standard procedure in the study of finite size effects in a second order phase transition is to use the scaling hypothesis: there is a single relevant length scale, $\xi$, which diverges at the critical point\ \cite{huang1987}, $\xi\sim (\gamma-\gamma_c)^{-\nu}$ where $\nu$ is a critical exponent and $\gamma_c$ the value of $\gamma$ for the transition. A different quantity displaying a non-analytical behaviour close to the critical point is then presumed to behave as $A(\gamma,L)\propto \hat{f}(\xi/L)$, where we assumed that $A$ does not have units of distance. In other words, the correlation length $\xi$ is the parameter that drives the transition. We can re-phrase this scaling ansatz as:  
\begin{equation}
A(L)=f(L^{1/\nu}(\gamma-\gamma_c)).\label{eq:scaling}
\end{equation}
where $f(x)$ is the so-called scaling function.
In the following we use this ansatz to characterize the two transitions in the RP model. 

\begin{figure}[t!]
\begin{centering}
\includegraphics[width=\columnwidth]{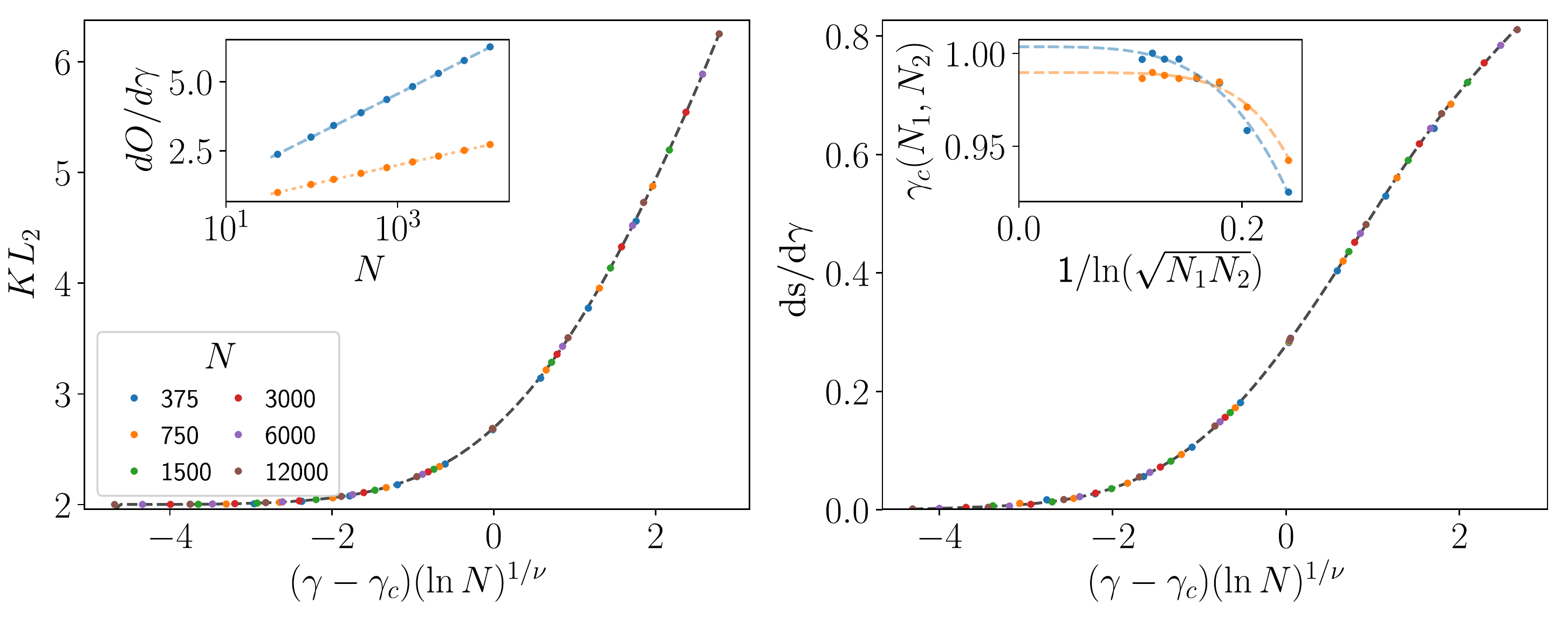}
\par\end{centering}
\vspace{0.2cm}
\caption{Left panel. $KL_2$ is plotted as a function of $x = \Delta\gamma(\ln N)^{1/\nu}$, with $\Delta\gamma=\gamma-1.00(4)$ and $\nu = 1.00(3)$, showing scaling collapse. In the inset, the slope of $KL_2$ and the derivative $\frac{ds}{d\gamma}$ close to the transition are plotted as a function of $N$ in log-log scale. Right panel. It is shown the scaling collapse of $\frac{ds}{d\gamma}$ as a function of $x = \Delta\gamma(\ln N)^{1/\nu}$, with $\Delta\gamma=\gamma-0.99(4)$ and $\nu = 1.04(5)$. Inset shows crossing points of the curves, for $KL_2$ (blue) and $\frac{ds}{d\gamma}$ (orange), for consecutive system sizes and as a function of $N$, in semi-logarithmic scale.
}\label{Fig_sc_g1}
\end{figure}

A first attempt to perform finite-size scaling using the system size as the dimension of the Hilbert space, $L=N$ in equation\ \ref{eq:scaling},  results in very small $\nu$ at both of the transitions. We have found that taking $L=\ln(N)$ in equation \ref{eq:scaling} produces better fittings than $L=N$, hence we use a scaling variable $x=(\ln N)^{1/\nu}(\gamma-\gamma_0)$. Similar scaling occurs for many-body system near a critical point, as the length of the system is related to the dimension of the Hilbert space as $L\sim\ln(N)$\ \cite{vojta2003}. Note also that a different scaling collapse had been achieved previously in a random regular graph  for moments of the wavefunctions \cite{garcia2017scaling}. There, quantities of the form $I_q\sim N^{D_q(q-1)}$ were scaled using two scaling variables, different at each side of the Anderson transition. Here instead we attempt scaling collapse of quantities that depends on $\log(N)$, as for example $KL_2= (\alpha_0-\alpha_1)\log(N)$, with a single scaling variable. 

In Fig.~\ref{Fig_sc_g1} we show the scaling collapse close to the transition with $\gamma_c \approx 1$ for both $KL_2$ (left panel) and ${\rm d}s/{\rm d}\gamma$ (right panel), being $s=S/\ln(N)$ the participation entropy divided by the logarithm of the matrix dimension. For each case we construct a cubic B-spline \cite{scipy} with 11 equidistant knots and minimise $\chi^2$. We restrict the range of $\gamma<1.35$ to avoid including corrections of order $(\gamma-\gamma_c)^2$ and the effects of the transition at $\gamma_c\approx 2$. We calculate error bars and asses the goodness of the scaling collapse with bootstrap techniques \cite{andrae2010}.  For $KL_2$ in left panel, we obtain a scaling collapse when $x= (\gamma-\gamma_c) (\ln N)^{1/\nu}$ for values of $\nu= 1.00(3)$ and $\gamma_c=1.00(4)$ \footnote{Error bars are one standard deviation estimated with bootstrap techniques.}. For ${\rm d}s/{\rm d}\gamma$ (right panel), parameters are estimated to be $\nu= 1.04(5)$ and $\gamma_c=0.99(4)$. Considering smaller system sizes in the scaling produces values of $\nu$ and $\gamma$ that drift slightly from the present ones due to finite size corrections. We find that this drift, which is studied in next paragraph, is negligible when $N\ge 375$ for these quantities.

From the data in the inset of left panel Fig.~\ref{Fig_sc_g1}, we can study how the crossing points in the quantities $KL_2$ (blue) and ${\rm d}s/{\rm d}\gamma$ (orange) drift for different sizes. The value of $\gamma$ at which each of these two quantities crosses for two consecutive system sizes, $N_1$ and $ N_2$, is plotted as a function of the inverse of the logarithm of their geometrical mean, $1/\ln( \sqrt{N_1 N_2})$. The average of the extrapolated critical points is $\lim_{N\to\infty}\gamma_c = 0.990(14)$.

We now use a different method to compute $\nu$ so we can check the consistency of the previous results. The inset in right panel of Fig.~\ref{Fig_sc_g1} shows the slope of $KL_2$ (blue) and ${\rm d}S/{\rm d}\gamma$ (orange) at the value of $\gamma_c=1$. A fit to  $A (\ln N)^{1/\nu} (1+ B/(\ln N)^{y_1})$ results in an average $\nu = 0.88(9)$. Thus, these estimates of critical parameters from the slopes of $KL_2$ and $ds/d\gamma$ are in good agreement with the ones from the data collapse discussed in previous paragraph.

 \begin{figure}[t!]
\begin{centering}
 \includegraphics[width=1.0\columnwidth]{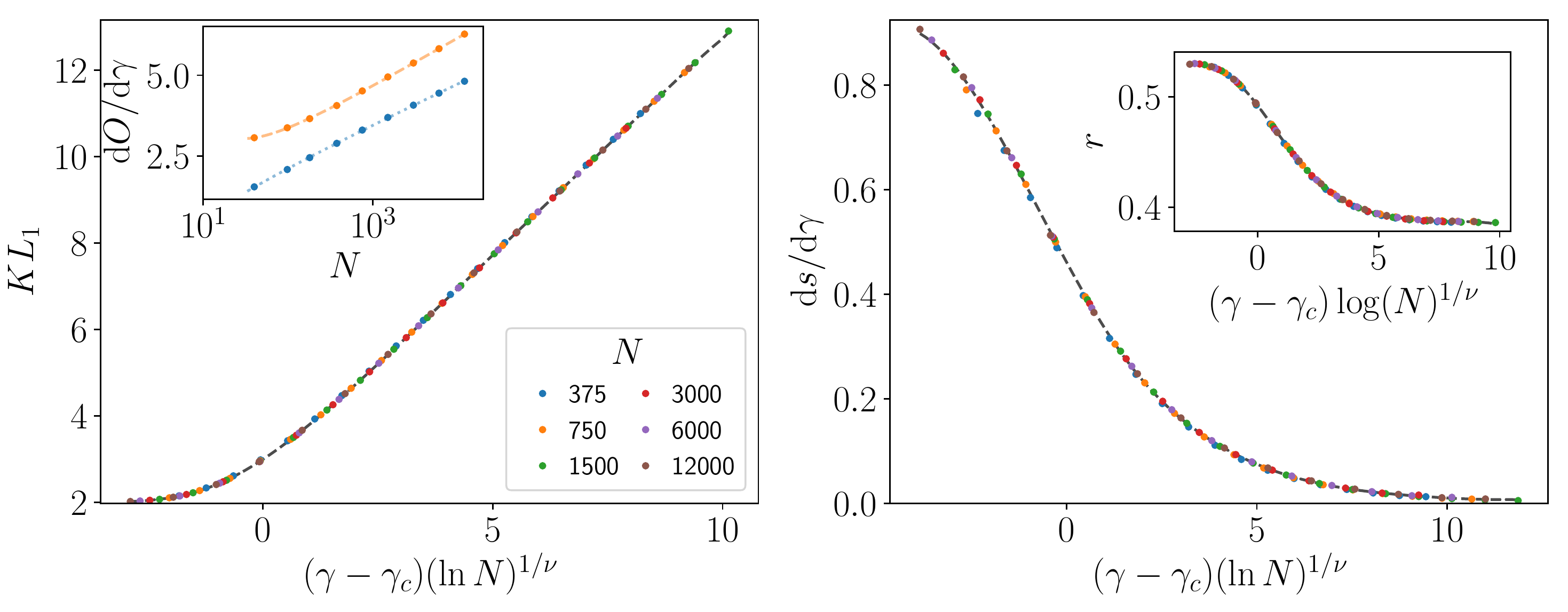}
\par\end{centering}
\vspace{0.2cm}
\caption{Left panel. Scaling collapse of $KL_1$ is plotted as a function of $x = \Delta\gamma(\ln N)^{1/\nu}$, with $\Delta\gamma=\gamma-2.009(5)$ and $\nu = 1.003(6)$. Inset shows the slope of $KL_1$ and derivative $\frac{ds}{d\gamma}$, where  $s=S/\ln(N)$, close to the transition as a function of $N$, in semi-log scale. Right panel. Scaling collapse of ${\rm d}s/{\rm d}\gamma$ as a function of $x = \Delta\gamma(\ln N)^{1/\nu}$, with $\Delta\gamma=\gamma-2.04(2)$ and $\nu = 0.91(7)$. In the inset, the $r$-statistic is shown as a function of the scaling variable, showing scaling collapse, where $\gamma_c=2.011(7)$ and $\nu=1.003(3)$. 
}\label{Fig_sc_g2}
\end{figure}

In Fig.~\ref{Fig_sc_g2} we study the transition close to $\gamma_c = 2$ for $KL_1$ (left panel) and ${\rm d}s/{\rm d}\gamma$ (right panel), where $s=S/\ln(N)$. As in the previous paragraph, we construct a cubic B-spline with 11 equidistant knots and minimise $\chi^2$, restricting the range in $\gamma>1.65$. For $KL_1$ (left panel), with $x = \Delta\gamma(\ln N)^{1/\nu}$ we obtain scaling collapse when critical values are $\nu= 1.003(6)$ and $\gamma_c=2.009(10)$. For ${\rm d}s/{\rm d}\gamma$ (right panel), scaling collapse is achieved for values $\nu= 0.91(7)$ and $\gamma_c=2.04(2)$. Critical values are thus consistent with $\nu=1$ and $\gamma_c=2$, although there is a small disagreement if one takes into account this estimate of the error bars. This small disagreement is due to the fact that finite size effects in this transition, at $\gamma=2$, are somewhat larger than in the previous studied case $\gamma=1$. We have performed the analysis of the drift of the crossing points (\emph{data not shown}), similar to the one in the inset of right panel of figure \ref{Fig_sc_g1}, which gives the location of the critical point at $\lim_{N\to\infty}\gamma_c = 1.97(3)$, in agreement with theoretical expectations. In addition, the slopes of the quantities mentioned before near the critical point are shown in inset of the left panel in Fig.~\ref{Fig_sc_g1}. These slopes grow with $(\ln N)^{1/\nu}$, where the two values of $\nu$ are compatible between them and with an average $\nu \approx 0.95(4)$.  

In previous discussion, we have presented the scaling collapse of quantities computed from the probability distribution of the eigenfunctions. In the thermodynamic limit, they are of the form $f(D_q)\log(N) $ where $f$ has linear dependence on the fractal dimensions $D_q$. However, we can attempt to get scaling collapse of other general quantities as the r-statistic. From Fig.\ \ref{Fig_sev}, we expect to get a non-trivial scaling collapse close to $\gamma=2$ (see section\ \ref{sec:cha_energies}). This is shown in the inset of left panel of figure \ref{Fig_sc_g2}. The parameters, obtained in a similar fashion as before, are $\gamma_c=2.011(7)$ and $\nu=1.003(3)$. This is another evidence of the general validity of the scaling ansatz, Eq.\ \ref{eq:scaling} with $L=\log(N)$, and it supports the scaling hypothesis. 

The results are fully consistent with scaling variables $x=(\gamma-\gamma_c)\ln(N)$ at both transition. Similar logarithmic scaling was obtained at Anderson localization transition for large space dimension\ \cite{tarquini2017,garcia2007dimensional}. The scaling variable near this type of metal-insulator transitions is given by  $x=(p-p_c)N^{\frac{1}{\nu d}}$, where the number of sites and linear size are related via the dimension as $N=L^d$. This is consistent with a logarithmic scaling $x\sim \log(N)$ in the limit of large $d$\ \cite{tarquini2017}. Note that the RP Hamiltonian contains hopping (non-diagonal terms in $H$) between any two site of the lattice. These all-to-all couplings make this model similar to a single-particle hopping in an infinite dimensional lattice.

\section{Summary}

We have used quantities based on the probability distribution of eigenenergies and eigenstates to characterize the three different phases of the RP model: ergodic, extended non-ergodic and localized. First, we have argued that the extended non-ergodic regime is chaotic as its eigenstates exhibit level repulsion. We have seen that distribution of level spacing -- one of the main tools used in the field of quantum chaos -- captures the properties of the non-ergodic extended to localized transition. However, it cannot  distinguish between non-ergodic and ergodic extended states. To overcome this limitation, we have used several quantities constructed from the eigenstates of the RP model to provide a full characterization of its phase diagram and phase transitions, including the ergodic to non-ergodic extended one at $\gamma=1$. 

We have performed a finite size analysis around the two critical points. The result shows that the scaling hypothesis, taking the logarithm of matrix dimension as the size of the system, is obeyed and we have obtained a critical exponent $\nu=1$ at both of the transitions. In other words, there is a single quantity that controls the divergence at each of the critical points.  These results evidence that the RP model posses a non-analytical behaviour similar to the one in a standard second order phase transition. We expect that these tools can be used in the study of non-ergodic extended to ergodic transitions in other models.

This picture of a second-order phase transition may not be universal. A first-order one was reported in a random-regular graph \cite{Al2016}. The reason of the differences regarding the order of the transitions may be rooted in the multifractal spectrum.  Indeed, the RP model exhibits a simple multifractal spectrum, similar to a fractal, while  the extended non-ergodic phase of random-regular graph may display stronger multifractal properties. 

Finally, we would like to comment on the possible implication of these results in the area of quantum information. The Hamiltonians that can be used in quantum annealing to solve hard problems may exhibit some of the properties of the RP model, as we explained in the introduction. The discussion of first versus second order phase transitions is important in the case that these Hamiltonians contains a low temperature phase of non-ergodic extended states. Quantum annealing would probably not  work -- adiabaticity condition would require an exponentially long time in system size -- in the case of a first order phase transition, while a second order phase transition only requires polynomial long times\ \cite{laumann2012quantum}. Second order phase transitions, as the one shown here for the RP model, would be much more benign for models of quantum annealing than first order ones, as reported for random regular graphs.

\section{Acknowledgments}

We thank G. Biroli for pointing out the importance of finite size effects in the distribution of level spacings. Financial support by Fundacion General CISC (Programa Comfuturo) is acknowledged. The authors acknowledge the computer resources and technical assistance provided by the Centro de Supercomputacin y Visualizacion de Madrid(CeSViMa) and by the cluster Trueno of the CSIC. PS thanks Fundaci\'on S\'eneca grant 19907/GERM/15.

\section*{Bibliography}
\bibliography{./MBLandNonequilibrium}

\end{document}